\documentclass[12pt,twoside,a4paper]{article}

  \usepackage{graphicx}
  \usepackage{epstopdf}
  \usepackage{calc}
  \usepackage{amssymb}
  \usepackage{amsmath}
  \usepackage{makeidx}
  \usepackage{footmisc}
  \usepackage{minitoc}
  \usepackage{longtable}
  \usepackage{fancyhdr}

  \newlength{\dinwidth} \newlength{\dinmargin}
  \mtcsetfont{secttoc}{subsection}{\normalfont}
\newcommand{\beq}[1]{\begin{equation}\label{#1}}
\newcommand{\eeq}{\end{equation}}
\newcommand{\bea}[1]{\begin{eqnarray}\label{#1}}
\newcommand{\eea}{\end{eqnarray}}

\newcommand{\Fi}[1]{Figure~\ref{#1}}

  \def\Journal#1#2#3#4{{#1}~{\bf #2}, #3 (#4)}

  \def\NP{Nucl.\ Phys.}
  \def\PL{Phys.\ Lett.}

  \def\CPC{Comput.\ Phys.\ Commun.}

\newcommand{\qcdnum}{\mbox{\sc qcdnum}}
\newcommand{\Qcdnum}{\mbox{\sc Qcdnum}}

\newcommand{\apfel}{\mbox{\sc apfel}}

\newcommand{\ceeF}{\ensuremath{C_{\rm F}}}
\newcommand{\teeR}{\ensuremath{T_{\rm R}}}
\newcommand{\ceeG}{\ensuremath{C_{\rm G}}}
\newcommand{\pee}[2]{P_{\rm #1}^{\rm (#2)}}
\newcommand{\peehat}[2]{\hat{P}_{\rm #1}^{\rm (#2)}}

\newcommand{\pa}{\partial}

\newcommand{\ms}{\ensuremath{\mu^2}}

\newcommand{\enef}{\ensuremath{n_{\rm f}}}

\newcommand{\as}{\ensuremath{\alpha_{\rm s}}}
\newcommand{\astpi}{\ensuremath{\frac{\as}{2\pi}}}
\newcommand{\asubs}{\ensuremath{a_{\rm s}}}

\newcommand{\qbar}{\ensuremath{\bar{q}}}

\newcommand{\de}{\delta}

\newcommand{\half}{\mbox{$\frac{1}{2}$}}     
\newcommand{\mbfrac}[2]{\mbox{$\frac{#1}{#2}$}}

\newcommand{\der}{\ensuremath{{\rm d}}}
\newcommand{\ve}[1]{\ensuremath{\boldsymbol{#1}}}

\newcommand{\xtt}{\texttt}

\newcommand{\mbstrut}[1]{\rule{0mm}{#1}}

 {\begin{list}{}{
  \settowidth{\labelwidth}{#1}%
  \setlength{\leftmargin}{\labelwidth+\labelsep}%
  \setlength{\itemsep}{#2}}}
 {\end{list}}

\title{
  \Large \bf Erratum for the time-like evolution in QCDNUM
}

\author{
  M. Botje\thanks{ 
  email m.botje@nikhef.nl}\\
  Nikhef, Science Park, Amsterdam, the Netherlands
}

\date{February 29, 2016}

\begin{document}

 \maketitle                        
 
\begin{abstract}
\noindent
A recent comparison of the evolution programs \qcdnum\ and \apfel\
showed a discrepancy in the time-like evolution of the
singlet fragmentation
function at~NLO. It was found
that the splitting functions of this evolution were wrongly assigned
in \qcdnum, and also that the fragmentation functions were not correctly
matched at the flavour thresholds.
These errors are corrected in a new release of the program.
\end{abstract}

\section{Introduction \label{se:Introduction}}
\Qcdnum~\cite{ref:qcdnum} is a fast QCD evolution program that can evolve
parton densities (space-like evolution) and fragmentation
functions (time-like evolution). 
Up to NLO, the evolution kernels are taken
from publications by Furmanski and Petronzio for the
flavour non-singlet~\cite{ref:fpns} and singlet 
evolutions~\cite{ref:fpsi}.\footnote{Well known misprints
in~\cite{ref:fpsi} can be found
in a footnote of~\cite{ref:qcdnum} and are corrected for. 
}

A recent
comparison~\cite{ref:valerio} of \qcdnum\ and the evolution program
\apfel~\cite{ref:apfel} has shown very good agreement between the codes,
except for the singlet evolution of fragmentation functions
at NLO. This is because \qcdnum\ used a NLO time-like 
splitting function matrix
in the index notation of~\cite{ref:fpsi}, instead of properly taking
its transpose. It also appeared that
the fragmentation functions were not correctly matched at the
flavour thresholds when running the evolution in the variable flavour
number~scheme at NLO.

The transposed matrix is implemented
in the new release \xtt{17-00/07}
of \qcdnum, together with the NLO threshold matching of the
fragmentation functions as described
in~\cite{ref:matching}.\footnote{
The errors are also fixed in
the beta releases version \xtt{17-01/12} and higher. All current 
\qcdnum\ releases can be downloaded from
\xtt{http://www.nikhef.nl/user/h24/qcdnum}.}

In \Fi{fig:apfel} 
%
\begin{figure}[tbh]
\begin{center}
\includegraphics[width=1.0\linewidth]{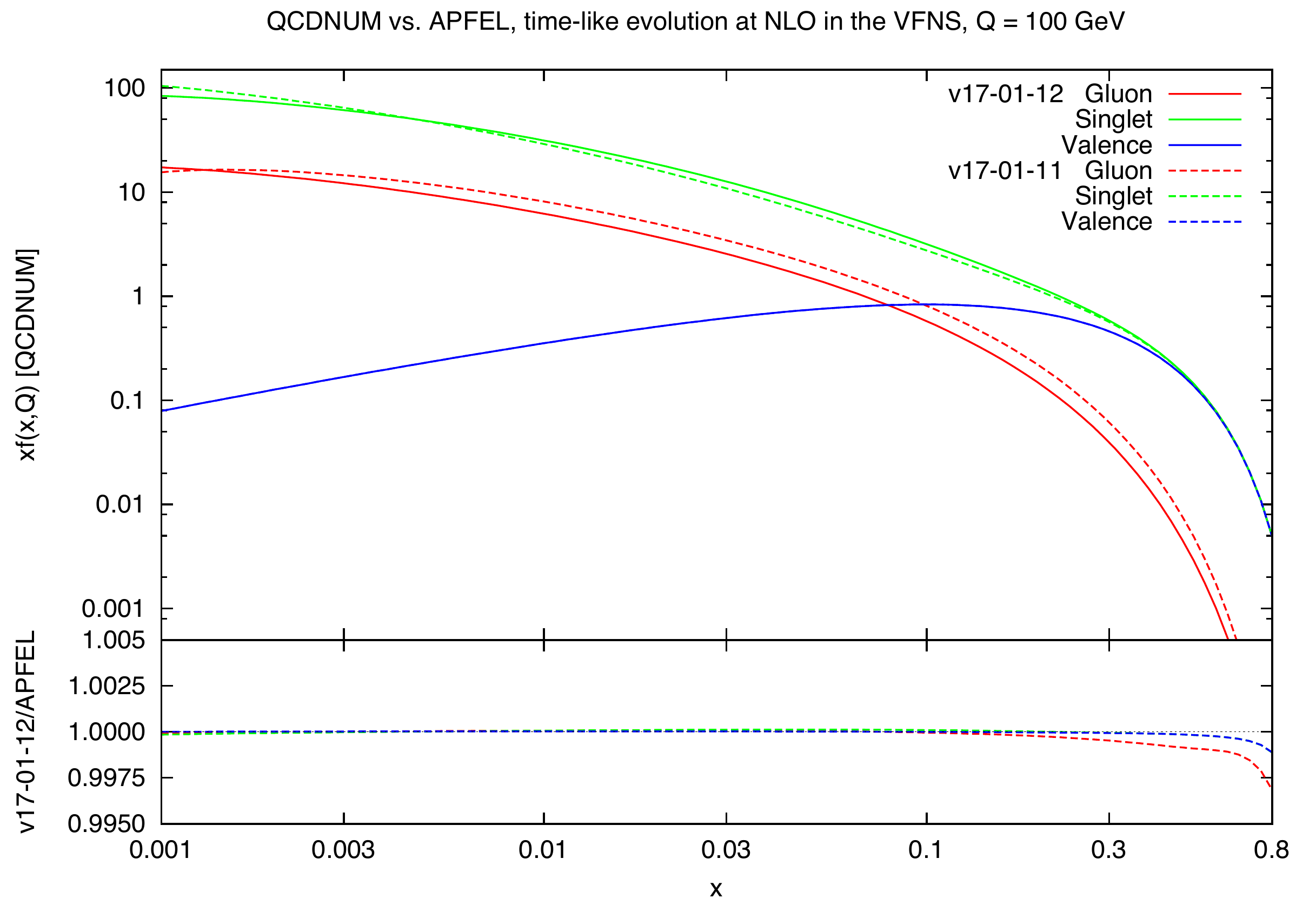}
\end{center}
\caption{\footnotesize 
  Time-like evolution at NLO in the variable
  flavour number scheme of 
  gluon, singlet and valence distributions
  to a scale of $\mu = 100$~GeV using an uncorrected (dashed curves) and 
  corrected version of \qcdnum\ (full curves). The lower panel shows
  the ratio of distributions
  evolved with the corrected \qcdnum\ version and \apfel.
  }
\label{fig:apfel} 
\end{figure}
%
we show the time-like evolution at NLO in the
variable flavour number scheme of the gluon, singlet and
valence distributions up to a scale of $\mu = 100$~GeV
with old (dashed curves)
and new versions of \qcdnum\ (full curves). There are sizeable differences
except in the valence evolution which is not affected
by the error in the splitting function matrix since it is a non-singlet.
In the lower panel of the plot is shown the comparison of the new \qcdnum\ version with \apfel.
It is seen that, after the correction in \qcdnum, the agreement between
the two evolution programs is excellent.

To clarify the index notation, we present in the
next section the splitting function
matrices that are currently implemented in \qcdnum.

\section{Singlet evolution  \label{se:evolution}}
We write the singlet evolution (coupled to the gluon)
in matrix notation as
 \[
 \frac{\pa \ve{V}}{\pa \ln \ms} = \ve{M} \otimes \ve{V} \quad
 \text{with} \quad \ve{V} = \begin{pmatrix}F \\ G \end{pmatrix}
 \quad \text{and } \ve{M} =
 \begin{pmatrix} P_{\rm qq} & P_{\rm qg} \\ 
                 P_{\rm gq} & P_{\rm gg} \end{pmatrix}.
 \]
 Here the symbol $\otimes$ denotes the Mellin convolution
 \[
   [f \otimes g](x) = \int_x^1 \frac{\der z}{z} f\left(\frac{x}{z}\right)
    g(z).
 \]
 For space-like evolution $G$ is the gluon density and $F$ is the
 quark singlet density
 $\sum_{i=1}^{\enef} (q_i + \qbar_i)$ 
 where $q_i$ ($\qbar_i$) is the (anti)quark number density of flavour
 $i$ in the proton and \enef\ is the number of active flavours.
 For time-like evolution $G$ and $F$ stand for the
 corresponding fragmentation functions.
  
 Below we will be only concerned with a splitting function
 expansion up to NLO,
 \[ 
 \ve{M} = \asubs\; \ve{M}^{(0)}  + \asubs^2\; \ve{M}^{(1)}
 \quad \text{with} \quad \asubs \equiv \astpi.
 \]
 The following four functions are defined in \cite{ref:fpsi}
 \[
 \begin{array}{ll}
 p_{\rm FF} = (1+x^2)/(1-x)   \qquad & p_{\rm GF} = x^2 + (1-x)^2 \\
 p_{\rm FG} = [1 + (1-x)^2]/x \qquad & p_{\rm GG} = 1/(1-x)+1/x-2+x-x^2
 \mbstrut{1.3em}.\end{array}
 \]
 The four LO splitting functions are then written as
 \[
 \begin{array}{ll}
 \pee{FF}{0} = \ceeF\; [p_{\rm FF} ]_+  \qquad &
 \pee{GF}{0} = 2 \teeR \enef\; p_{\rm GF} \\
 \pee{FG}{0} = \ceeF\; p_{\rm FG}   \qquad &
 \pee{GG}{0} = 2 \ceeG x^{-1} [ x p_{\rm GG} ]_+ - \mbfrac{2}{3} \teeR
 \enef\; \de(1-x) \mbstrut{1.6em} \end{array}
 \]
 with the colour factors and the regularisation prescription given by
 \[
 \ceeF = \mbfrac{4}{3},\quad \ceeG = 3,\quad \teeR = \half \quad \text{and}
 \quad [f(x)]_+ \equiv f(x) - \de(1-x) \int_0^1 f(y) \der y.
 \]   
 The NLO splitting functions for space-like (S) and time-like (T) processes
 are
\bea{}
 \pee{FF}{1,U} & = & \peehat{FF}{1,U} - \de(1-x) \int_0^1 \der x\; x
 \left[\peehat{FF}{1,T} + \peehat{FG}{1,T} \right] \nonumber  \\
 \pee{GF}{1,U} & = & \peehat{GF}{1,U}   \nonumber  \\
 \pee{FG}{1,U} & = & \peehat{FG}{1,U}   \nonumber \mbstrut{1.5em} \\
 \pee{GG}{1,U} & = & \peehat{GG}{1,U} - \de(1-x) \int_0^1 \der x\; x
 \left[\peehat{GG}{1,T} + \peehat{GF}{1,T} \right], \nonumber
\eea
where U = \{S,T\}. The functions $\peehat{AB}{1,U}$ are
given in Eqs. (11) and (12) of \cite{ref:fpsi}.\footnote{
Modulo the misprint in $\peehat{FF}{1,T}$ which does not affect
the colour factors.}

Because the authors of~\cite{ref:fpsi} do not clearly define their
index notation (hence the confusion), we identify the splitting functions
not by their indices but, instead,
by their overall colour factors which should be the same at LO
and NLO.

For space-like evolution the colour factors of $P_{\rm qg}$ 
and $P_{\rm gq}$ are
$2 \teeR \enef$ and $\ceeF$ while those for
time-like evolution are
$2 \ceeF \enef$ and $\teeR$, respectively. 

Identifying the splitting functions by these factors we obtain
the LO and NLO space-like
evolution matrices (note that these were always correctly
implemented in \qcdnum):
\beq{eq:spacelike}
 \ve{M}^{\rm (0,S)} = \begin{pmatrix}
 \pee{FF}{0} & \pee{GF}{0} \\
 \pee{FG}{0} & \pee{GG}{0} \mbstrut{1.7em} \end{pmatrix}, \qquad\quad
 \ve{M}^{\rm (1,S)} = \begin{pmatrix}
 \pee{FF}{1,S} & \pee{GF}{1,S} \\
 \pee{FG}{1,S} & \pee{GG}{1,S} \mbstrut{1.7em} \end{pmatrix}.
\eeq
  
It is well known that the LO time-like matrix is
the transpose of the space-like matrix~\cite{ref:nasonwebber}.
To get the same colour factors at NLO it can be seen from inspection
of Eq. (12) in~\cite{ref:fpsi} that also the NLO matrix must be 
transposed.
Accounting for factors $2\enef$, we thus have
\beq{eq:timelike}
 \ve{M}^{\rm (0,T)} = \begin{pmatrix}
 \pee{FF}{0} & 2\enef \pee{FG}{0} \\
 \frac{1}{2\enef}\pee{GF}{0} & \pee{GG}{0} \mbstrut{1.7em} \end{pmatrix},
 \qquad\quad
 \ve{M}^{\rm (1,T)} = \begin{pmatrix}
 \pee{FF}{1,T} & 2\enef \pee{FG}{1,T} \\
 \frac{1}{2\enef}\pee{GF}{1,T} & \pee{GG}{1,T} \mbstrut{1.7em}   
 \end{pmatrix}.
\eeq
The mistake made in the previous \qcdnum\ releases is that the
NLO time-like matrix~$\ve{M}^{\rm (1,T)}$ 
was \emph{not} transposed, contrary to what is
done in \apfel\ \cite{ref:apfeltimelike}.

\section{Acknowledgements}
I am grateful to V. Bertone for spotting the problem, for running
several \qcdnum\ versus \apfel\ comparisons to locate the error,
and for providing the plot. I thank
A. Vogt for useful discussions. Finally, my apologies to those who
have used previous versions of \qcdnum\ to evolve their singlet
fragmentation functions at NLO.



\end{document}